\documentclass[preprint,12pt]{aastex} 
\usepackage{cyr}
\usepackage{epsf}
\usepackage[reqno]{amsmath}
\usepackage{amsfonts}
\usepackage{amssymb}
\usepackage{bm}
\usepackage{hyperref}
\tightenlines

\def\bes{{}^7\rm{Be}}
\def\best{{}^7\rm{Be}^{2+}}
\def\besth{{}^7\rm{Be}^{3+}}
\def\besf{{}^7\rm{Be}^{4+}}
\def\lis{{}^7\rm{Li}}
\sloppypar

\begin{document}
\baselineskip 21pt

\title{\bf Time of primordial ${}^7$Be conversion into ${}^7$Li, energy
  release and doublet of narrow cosmological neutrino lines}

\author{Rishi Khatri$^{*1}$}
\email{${}^*$khatri@mpa-garching.mpg.de}
\and 
\author{Rashid A. Sunyaev$^{1,2}$}

\affil{
{\it Max Planck Institut f\"{u}r Astrophysik, Karl-Schwarzschild-Str. 1,
  85741, Garching, Germany}$^1$\\ 
{\it Space Research Institute, Russian Academy of Sciences, Profsoyuznaya 84/32, 117997 Moscow, Russia}$^2$}

\vspace{2mm}
\begin{abstract}
 One of the important light elements created during the big bang
  nucleosynthesis is ${}^7$Be  which then decays to ${}^7$Li by electron
  capture when recombination becomes effective but well before the Saha equilibrium recombination is reached. This means that ${}^7$Be should wait until its recombination epoch
  even though the half-life of the hydrogenic beryllium atom  is only $106.4$ days. We calculate when 
  the conversion from primordial ${}^7$Be to ${}^7$Li occurs taking into
  account the population of the hyperfine structure sublevels and solving
  the kinetic equations for recombination, photoionization and conversion
  rate. We also calculate the energies and the spectrum of narrow neutrino doublet lines resulting from $\bes$ decay.
\end{abstract}
\hspace{4 pt}\\
\noindent PACS: 98.80.Ft,98.70.Vc\\
Key words: \emph{cosmic microwave background,  cosmic neutrino
  background, big bang nucleosynthesis, early Universe}
\pagebreak

\section{Introduction}
The first major event  in the Universe after the
big bang and the inflation stage is the creation of light elements
\citep{wfh67} 
 including D, He-3 and Li, which are of importance for our everyday life or considered as energy source on the Earth in the future.  The combination of neutron and proton
to form deuterium is the first step in the nuclear reaction chain. The big
bang nucleosynthesis (BBN) starts
when the temperature of the Universe is about  $T\sim 0.1$MeV (redshift $z\sim
4\times 
10^8$) and  the deuterium
destroying high
energy photons have become rare. BBN  proceeds swiftly after this deuterium
bottleneck is overcome and is over by $T\sim 0.01$MeV. The next major event
in the Universe is that of the recombination of primordial plasma
(\cite{zks68};\cite{peebles68}) to form
neutral atoms starting with helium III to helium II recombination  at $4500\lesssim z \lesssim 7000$ and
ending with hydrogen recombination at $800 \lesssim z \lesssim 1800$. In this paper we will
explore a unique connection between these two major events, the
recombination of beryllium ${}^7$Be and its decay to lithium $\lis$ and
in the process also mention a doublet of narrow lines in the cosmic neutrino
background. This unique event brings together the MeV scale nuclear
physics of nucleosynthesis and eV scale atomic physics of
recombination. We will also find that the very low energy atomic physics of
hyperfine transitions in $\bes$ has a surprisingly strong influence on the
decay properties of $\bes$ usually associated with  MeV scale physics. This
problem of beryllium to lithium conversion is one of the last
 problems  not investigated in detail so far 
connected with the primordial nucleosynthesis and the process of
recombination of ions and electrons in the radiation field of the cosmic
microwave background (CMB). This problem is very
attractive because it demonstrates a very rare case in astrophysics when
the atomic processes and even the populations of the hyperfine structure
influences nuclear processes.

The cosmological parameters used for numerical calculations are
Hubble constant $H_0=73\rm{km/s/Mpc}$, matter fraction $\Omega_m=0.24$,
CMB temperature $T_{CMB}=2.726K$, baryon
fraction 
$\Omega_b=0.043$,  cosmological constant $\Omega_{\Lambda}=0.76$.

\section{Decay of ${}^7\rm{Be}$}
At the end of BBN the abundances of light elements relative to hydrogen
are \citep{serpico2004}: $X_{{}^4\rm{He}}=6\times 10^{-2}$, 
$X_{{}^2\rm{H}}=2.5\times 10^{-5}$, $X_{{}^3\rm{He}}=10^{-5}$,
$X_{{}^3\rm{H}}\sim 10^{-7}$, $X_{{}^7\rm{Be}}\sim 10^{-10}$,
$X_{{}^7\rm{Li}}\sim 10^{-11}$, $X_{{}^6\rm{Li}}\sim 10^{-14}$ and
$X_{\rm{M}} \lesssim  10^{-15}$, where $X_{\rm{M}}$ is the abundance of
all heavier elements, $X_i=n_i/n_{\rm{H}}$, $n_i$ is the abundance of
species $i$ and $n_{\rm{H}}$ is the number density of hydrogen nuclei.
Note that about $\sim 90\%$ of the primordial ${}^7\rm{Li}$ comes from
${}^7\rm{Be}$ and only $\sim 10\%$ of ${}^7\rm{Li}$  is produced directly
in  BBN, this division is however sensitive to the baryon to photon ratio
$\eta$ \citep{bbnreview}. 

${}^7\rm{Be}$ in its fully ionized state in the early Universe is
stable. Fully ionized beryllium can capture electrons from the plasma at
the high densities in the stellar interiors \citep{bahcall}. However the
electron capture rate for the fully ionized beryllium from the
plasma is completely negligible in the early
Universe due to the low electron density.  Once
recombined with an electron beryllium decays by
electron capture through the following reactions 
\begin{align}
{}^7\rm{Be}+ e^- \rightarrow {}^7\rm{Li}+ \nu_e \label{r1}\\
{}^7\rm{Be}+ e^- \rightarrow {}^7\rm{Li}^{\ast}+ \nu_e,\hspace{4 pt}
{}^7\rm{Li}^{\ast}\rightarrow {}^7\rm{Li} + \gamma,\label{r2}
\end{align}
where ${}^7\rm{Li}^{\ast}$ is the excited state of lithium nucleus with
energy $477.6$KeV above the ground state. 
  The Q-value of the reaction is
$861.8$KeV \citep{tilley2002}. The laboratory value for the half life of 
$\bes$ is 53.2 days. The branching ratio is $10.44\%$ for reaction \ref{r2}.
These laboratory values have been calculated and measured for neutral $\bes$ with four
electrons. 

The $\bes$ in the primordial plasma will however decay as soon as
it acquires a single electron. Therefore we are interested in hydrogen-like
$\besth$. The electrons in
the $2s$ shell of neutral $\bes$ have a $\sim 1\%$ effect on the decay rate as shown by
experiments involving $\rm{Be}^{2+}(\rm{OH_2})_4$ as well as other chemical
compositions \citep{exp1} and we
 will ignore the influence of 2s shell electrons in the discussions below.  Helium-like $\best$ has practically the same lifetime and
branching ratio as the neutral $\bes$. The change in the decay rate and the
branching ratio  from helium-like $\best$ to hydrogen-like $\besth$ depends strongly on the spin temperature of the hyperfine structure of  hydrogen-like $\besth$ ions.
 
This statement is easy to understand. We will make simple estimates of
decay rates based
on the calculations by \cite{becalc}. $\lis^{\ast}$ has a nuclear spin of
$I=1/2$ which is different from the nuclear spin of $I=3/2$ of $\lis$ and
$\bes$. The two initial states available for $\besth$ are with total angular
momentum $F^+=2$ and $F^-=1$. The two final states available to
$\lis^{\ast}+\nu_e$ are $F=0,1$. Thus
the angular momentum can only be conserved for the reaction \ref{r2} from
the initial state
$F^-=1$ of $\besth$ to the final state $F=1$ of $\lis^{\ast}$.  The reaction is
suppressed for the other initial hyperfine state with total angular momentum
$F^+=2$. Correspondingly for  reaction \ref{r1},  where the nuclear spin does
not change in the reaction, the two available final states have the same
total angular momentum as the initial state, $F=1,2$, and the reactions from both
the initial hyperfine states are possible.
 Millielectronvolt hyperfine splitting dictates the branching ratio for
 these two nuclear reactions.  In hydrogen-like $\besth$ ion the population
 levels of the two hyperfine states (or equivalently the spin temperature) are
 determined by the rates of radiative decay and pumping due to well known physical processes.

Our calculation is simplified because pumping by the CMB radiation field
and due to electron collisions is much faster than the decay rate and
cosmological time, i.e. the spin temperature should be close to the CMB
temperature $T$ equal at that time to the electron temperature. The energy
of the hyperfine splitting is much smaller than the temperature at that
time. This means that the spin temperature is very high compared to the
hyperfine splitting and the hyperfine structure sublevels are populated according to their statistical weights.

\begin{figure}[h]
\epsffile{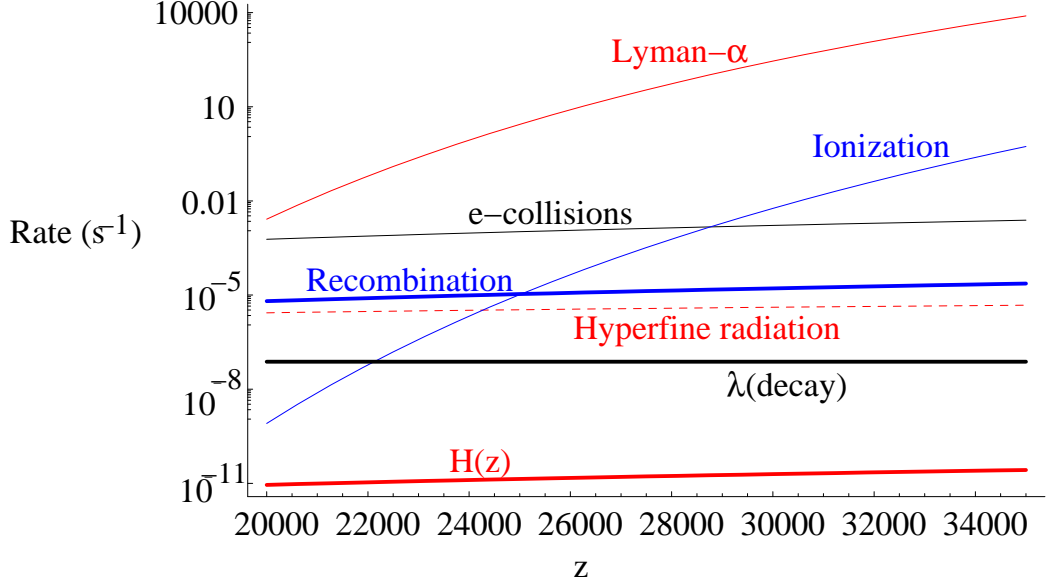}
\caption{Comparison of different processes influencing beryllium to lithium
decay. It is interesting to note that a typical hydrogenic beryllium atom
would have flipped its electron spin a billion times due to Lyman-$\alpha$
scattering and recombined and ionized $100$ times before finally being able
to decay.}
\label{fig1}
\end{figure}

The Lyman-$\alpha$ resonant scattering \citep{field,vars} with rate
$P_{\alpha} \sim 4\pi \int d\nu \sigma_{\nu}I_{\nu}/(h\nu) \sim
100\rm{s}^{-1}$ at $z\sim 30000$, spin changing collisions with electrons with rate
$\sim n_e \kappa_{10} \sim 10^{-4}\rm{s}^{-1}$ \citep{furn} and stimulated
emission and absorption of CMB at hyperfine transition with rate $\sim B_{21}I_{\nu}\sim
4\times 10^{-6}\rm{s}^{-1}$ are all faster than the decay rate $\lambda \sim
7\times 10^{-8}\rm{s}^{-1}$ and would make the spin temperature equal to the
matter/radiation temperature $T$. Above $\sigma_{\nu}$ is the absorption
cross section for Lyman$-\alpha$ photons including the line profile from Doppler broadening. $I_{\nu}$
is the background CMB intensity. It corresponds to Wien region of the black
body spectrum for the Lyman-$\alpha$
photons and Rayleigh-Jeans region for the hyperfine transition
radiation.  $B_{21}$ is the Einstein B coefficient for stimulated emission. $n_e$ is the electron number density and $\kappa_{10}$ is the
spin change cross section.  $\kappa_{10}\propto Z^{-2}$ for hydrogenic
ions  \citep{amg89}, where $Z$ is the nuclear charge. There is also a mild
energy dependence of the cross section at temperatures of interest of $\sim
1/T^{1/2}$. The spin change cross section for hydrogen-electron
collisions  is of the
order $\sim 10^{-9}\rm{cm}^3\rm{s}^{-1}$ at $T\sim 10^5$K (\cite{smith};\cite{furn}) and thus for $\besth$ it will be
$\sim 10^{-10}\rm{cm}^3\rm{s}^{-1}$. 
Thus Lyman-$\alpha$ scattering, Ionization and recombination, Collisions with electrons  and stimulated emission and radiation of CMB and
maintains $T_{spin}=T$. These rates are plotted in Figure \ref{fig1} along
with the expansion rate defined by the Hubble parameter $H(z)$. It is interesting to note that a typical hydrogenic beryllium atom
would have flipped its electron spin a billion times due to Lyman-$\alpha$
scattering and recombined and ionized $100$ times before finally being able
to decay. The energy difference between
the two hyperfine states is $10^{-4}\rm{eV}<< T$. Thus the hyperfine states
would always be distributed according to their statistical
weights. 

\begin{figure}[h]
\epsffile{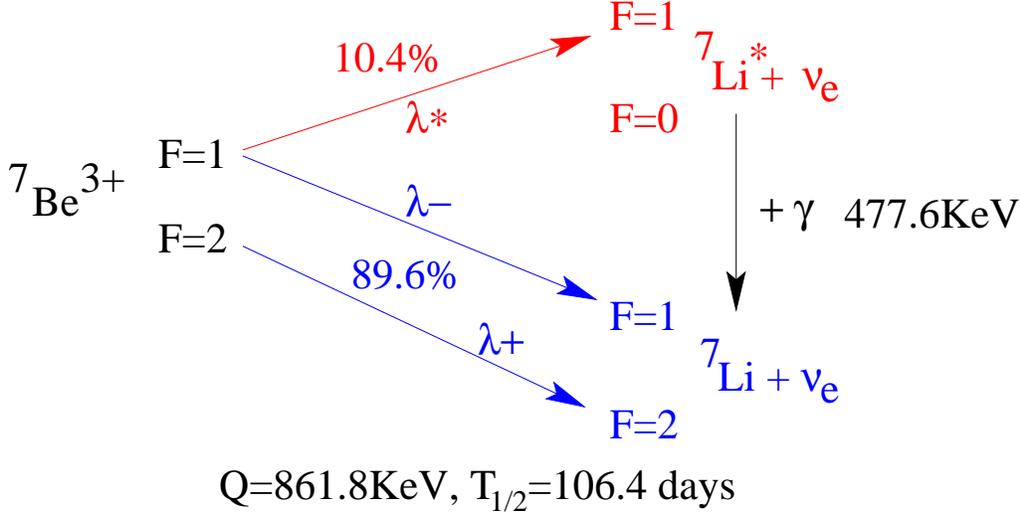}
\caption{Effect of hyperfine splitting of $\besth$ on the nuclear
  reactions. The half-life given above corresponds to the high redshift universe
  when the spin temperature is maintained at a high value by collisions
  with electrons. The half-life would be different if the  populations
  of hyperfine levels of hydrogenic $\besth$ are
  not distributed according to their statistical weights.}
\label{fig2}
\end{figure}

\begin{figure}[h]
\epsffile{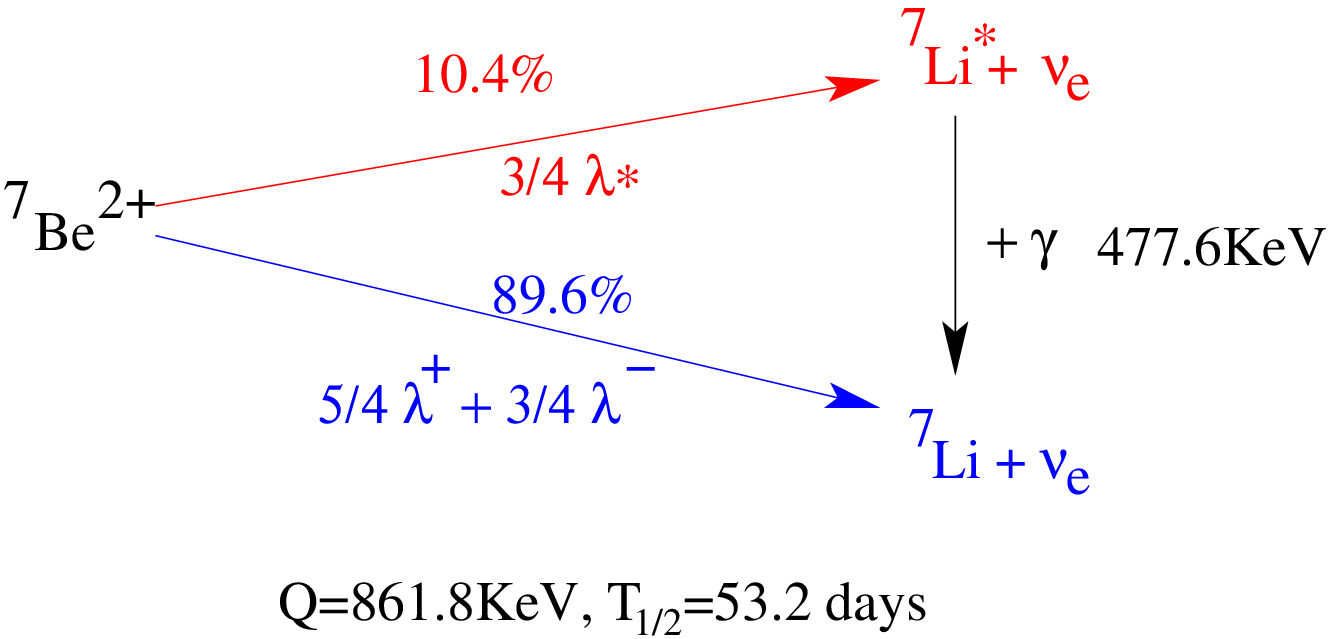}
\caption{Decay of helium-like $\best$.}
\label{fig3}
\end{figure}

In helium-like $\best$ (or Li-like and neutral $\bes$) spin directions of two $1s$
shell 
electrons is always opposite and the problem radically differs from the
problem in the hydrogen-like ion. Thus
there is no hyperfine splitting in the case of helium-like ions, but the relative
direction of the spins of the captured electron and nucleus is of great
importance.  We can write the probabilities of the two reactions in helium-like
ions in terms of the corresponding probabilities of the hydrogen-like ions
in the two hyperfine states using the recent theoretical calculations of \cite{becalc}
which have the support of  experiments involving hydrogen-like and helium-like ${}^{142}\rm{Pm}$ ions \citep{beexp}. Then we
can use these relations and the relative  populations of the hyperfine structure sub-levels in
the hydrogen-like  ions to find the formula for the probabilities of both channels of
interest. The reactions for hydrogen-like $\besth$ and helium-like $\best$ are summarized in Figures
\ref{fig2} and \ref{fig3}.

For reaction \ref{r1} the initial and final spin of the nuclei are same, $I=3/2$ and
the electron capture rates of $\best$ and $\besth$ are related by \citep{becalc}
\begin{equation}\label{eq3}
\lambda_{\best}=\left(\frac{2F^++1}{2I+1}\right)\lambda^+ + \left(\frac{2F^-+1}{2I+1}\right)\lambda^-,
\end{equation}
 where $\lambda^{\pm}$ are the decay rates from two
hyperfine states of $\besth$. If $f^{\pm}$ are the fraction of atoms in the
two hyperfine states, the net decay rate is given by 
\begin{align}\label{eq4}
\lambda_{\besth}&=f^+ \lambda^{+}
+f^- \lambda^{-} \nonumber\\
&=\left(\frac{2F^++1}{2(2I+1)}\right)\lambda^+ +
\left(\frac{2F^-+1}{2(2I+1)}\right)\lambda^-\nonumber\\
&=\lambda_{\best}/2,
\end{align}
where we used the fact that $(2F^++1)+(2F^-+1)=2(2I+1)$. The factor of two
can be understood as follows. In the helium-like atom both
channels corresponding to $\lambda^+$ and $\lambda^-$ are available
simultaneously which are added according to the statistical weights. For
the hydrogen-like atom also for high spin temperature we add the
probabilities of two channels according to the statistical weights. But for
the helium-like atom the capture rate gets multiplied by a factor of two
corresponding to the fact that any of the available electrons can be
captured through any of the available channels. Thus the factor of two is
the result of the fact that there are two electrons available for
helium-like atom and that for high spin temperature for the hydrogen-like
atom we add the capture rates
according to their statistical weights. This factor would, in general,
be different from two if the atoms in the hyperfine state were not
distributed according to their statistical weights, for example at low
temperatures comparable to the energy difference between the two states.
Since the statistically averaged rate coefficients relevant for our
calculation for the helium-like and hydrogen-like atoms
differ by a just factor of two, we do not need the individual rates for
$\lambda^+$ and $\lambda^-$ and can use the experimental result for the
helium-like/neutral $\bes$.

 For reaction \ref{r2}
however the $\lis^{\ast}$ has a spin of $1/2$ and only capture from $F^-$
state is allowed. The above formula is valid with $\lambda^+=0$ and for the
net
decay rate we have the same result,   $\lambda_{\besth}^{\ast}=\lambda_{\best}^{\ast}/2$. We are using ${}^{\ast}$ for the reaction rates of reaction
\ref{r2} while no ${}^{\ast}$ indicates the reaction rates for  reaction
\ref{r1}. Using the laboratory value of
$\lambda_{\rm{lab}}=\ln{2}/53.2\rm{days}$ for the total decay rate
of $\best$, we get net decay rate 
$\lambda_{\rm{bbn}}=0.5\lambda_{\rm{lab}}$ with the same branching ratio, 
 $10.4\%$ of the decays following reaction \ref{r2}. The half life  of
hydrogen-like $\besth$ with the hyperfine states distributed according to
their statistical weights is thus equal to 106.4 days. Coincidently this is the
same result we would have got from the simple considerations of the electron
density at the nucleus. We must emphasize here that the exact life time of
$\besth$ is not very important for us as long as it is much shorter than
the cosmological time at the epoch of beryllium to lithium conversion and
the duration of $\besth$ recombination. All
the width of the two neutrino lines  comes from the kinetics of recombination.
 In particular we have ignored the electron-electron interaction in the
 helium-like $\best$, the effect of which on our calculation of time of
 $\bes$  decay and the neutrino spectrum would be  $\sim few \%$.

\section{Recombination of ${}^7\rm{Be}$}
\begin{figure}[h]
\epsffile{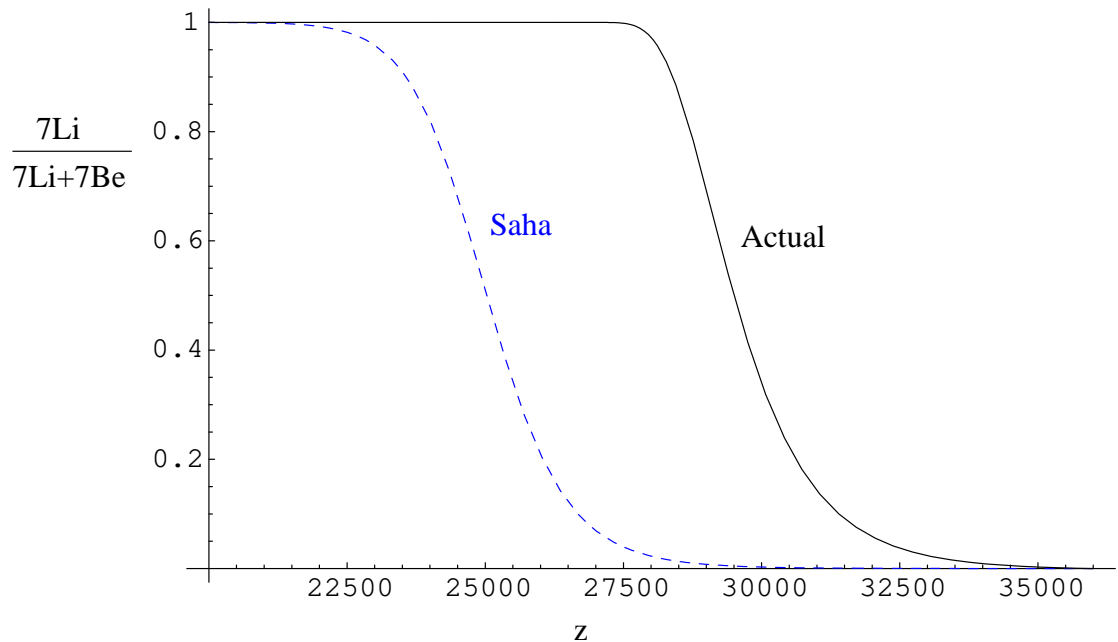}
\caption{Lithium number density as a fraction of total
  beryllium and lithium number density from decay of beryllium. For
  comparison thermal equilibrium results using Saha equation are also
  plotted.}
\label{fig4}
\end{figure}

Now  we can write down the kinetic equations for recombination of $\besf$.
\begin{align}
\frac{dX_{\besf}}{dz} & =
\frac{1}{H(z)(1+z)}\left[n_e(z)X_{\besf}\alpha_{\besf} -
  \beta_{\besth}X_{\besth}\right]\\
\frac{dX_{\besth}}{dz} & =
\frac{1}{H(z)(1+z)}\left[-n_e(z)X_{\besf}\alpha_{\besf} +
  \beta_{\besth}X_{\besth}+\lambda_{\rm{bbn}}X_{\besth}\right]\\
X_{\bes} & = X_{\lis}^{decay}+X_{\besf}+X_{\besth},
\end{align}
where $n_e(z)$ is the number density of electrons at redshift $z$, $H(z)$ is the Hubble
parameter, $\alpha_{\besf}$ is the total recombination coefficient
including recombination to the ground state. $X_{\lis}^{decay}$ is the lithium fraction coming from decay of
beryllium and excludes the  lithium produced during BBN. 
Because of the extremely low
number density of beryllium, the number of ionizing photons
released during direct recombination to the ground state is negligible
(about 10 orders of magnitude less)
compared to the ionizing photons already present in the background
radiation and can be neglected.  $\beta_{\besth}$ is the total ionization coefficient
 which is related to $\alpha_{\besf}$ by the condition that Saha equation
 must be satisfied in equilibrium. Thus 
\begin{equation}
\beta_{\besth}=\alpha_{\besf}
\frac{(2\pi
 m_e k_B T)^{3/2}}{(2\pi \hbar)^3}e^{\frac{-\chi_{Be}}{k_BT}}.
\end{equation}
Note that the 2-photon decay rate from $2s$ state of $\besth$ is $3.4\times
10^4 \rm{s}^{-1}$ \citep{goldman}. This is much faster than the same rate
for hydrogen $\sim 8 \rm{s}^{-1}$ due to the higher charge of the nucleus. More importantly it is much faster than the ionization rate
 $\beta \sim 6\times 10^{-3} \rm{s}^{-1}$ at $z \sim 30000$ and the electron capture rate
$\lambda_{bbn}=7.5\times 10^{-8}\rm{s}^{-1}$, $t_{1/2}=106.4$days. Thus $2s$ level  must be included in
  the total recombination coefficient. The number of Ly$\alpha$ photons
  produced would be of order $X_{\bes}$ and can be neglected. Thus
  equations for beryllium recombination are much simpler than that of
  hydrogen and helium recombination.
  $\alpha_{\besf}$ is the  total recombination coefficient including direct
  recombinations to the ground state and is
  given by \citep{ppb91}
\begin{equation}
\alpha_{\besf} =  10^{-13}Z\frac{at^b}{1+ct^d}\hspace{4 pt} \rm{cm}^3\rm{s}^{-1},
\end{equation}
where $t=(T/10^4K)/Z^2$, $a=5.596$, $b=-0.6038$, $c=0.3436$ and $d=0.4479$
and $Z$ is the nuclear charge..
Figure \ref{fig4} shows the result of integrating the recombination
equations for beryllium. For comparison the Saha equilibrium solution for
beryllium recombination is also shown. The beryllium to lithium conversion
occurs significantly earlier at $z=30000$ than the $z=25000$  value
predicted by the
Saha solution. The reason of the difference is connected with the short
decay time of recombined $\besth$. In the Saha equation we follow the
balance between the recombination and photoionization but a typical atom
has recombined and ionized many times even though the net recombination in
equilibrium may be small.  In reality
due to the decay of beryllium  on a time scale much shorter than the
cosmological time the equilibrium is
never established.

\section{Energy release and narrow doublet of neutrino lines}
\begin{figure}[h]
\epsffile{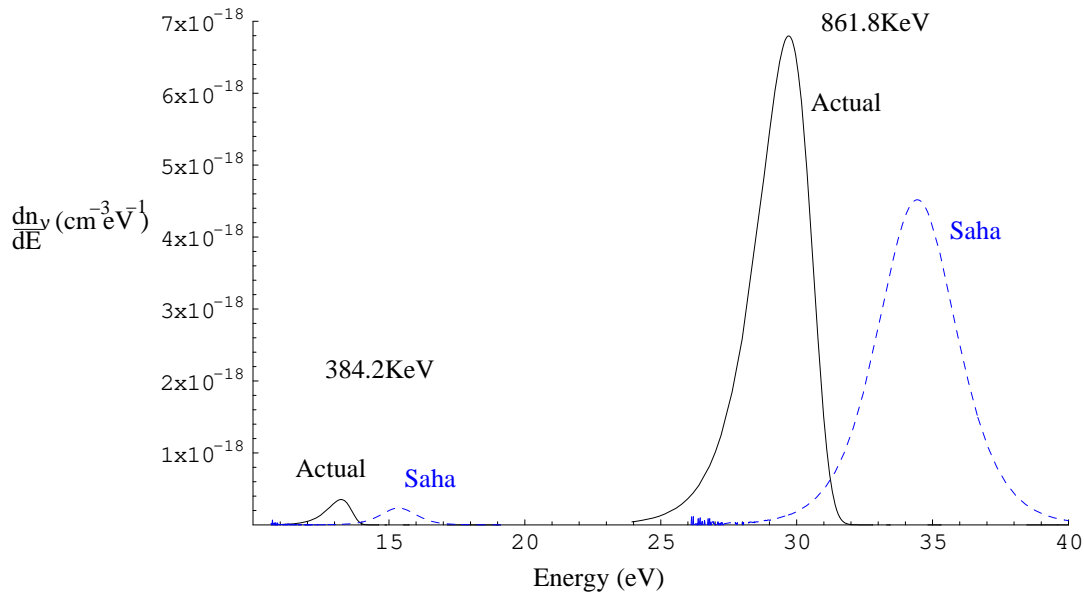}
\caption{Neutrino spectrum from beryllium decay in the early
  Universe. There are two lines corresponding to two electron capture decay
branches. For comparison equilibrium results (marked Saha) are also
plotted. The FWHM for high energy line is $2.3$eV with a central frequency of
$29.5$eV. For second line FWHM is $1$eV
central frequency is $13.1$eV. The line width at half maximum is $\delta
E/E =7.8\%$. The width and asymmetric line profile of these lines is defined by the
kinetics of recombination.}
\label{fig5}
\end{figure}

Every decay of beryllium to lithium is accompanied by emission of an
electron neutrino with $89.6\%$ of them having energy of $861.8$KeV and
nuclear recoil energy of $56.5$eV. $10.4\%$ of the decays go to excited
state of lithium, electron neutrino gets $384.2$KeV and nuclear recoil has
energy of only $11.2$eV which is slightly larger than the thermal energy
$T=7eV$ of the plasma at
redshift $z=30000$. The recoil velocity of $v/c=5.85\times 10^{-5}$ will result in a natural line width of $56$eV due
to Doppler shift for
the $477.6$KeV photon emitted when lithium relaxes to ground state almost
instantaneously ($T_{1/2}=73$fs). These
photons will down scatter by Compton scattering with electrons giving most of their energy
to the plasma due to recoil effect \citep{zs1969}. The energy transfer by Compton scattering will become
inefficient when the photons reach the critical energy of
$m_eH(z)/n_e\sigma_T|_{z=30000} \sim 80$eV when the energy transfer rate
becomes less than the expansion rate (\cite{bd90}). This will leave a small distortion in
the high energy part of CMB. The energy transfered to plasma will
additionally cause a $y-$type distortion of $\sim
(0.104E_{\gamma}/T)X_{\bes}\eta\sim 10^{-16}$. The neutrinos however will
free stream  to us and we can calculate the neutrino spectrum today. This
is plotted in Figure \ref{fig5} including both $861.8$KeV neutrinos and
$384.2$KeV neutrinos,
\begin{equation}
-\frac{dn_{\nu}}{dE}(E)=n_{H0}\left[\frac{0.896(1+z)^2}{861.8\rm{KeV}}\frac{dX_{\lis}^{decay}}{dz}|_{1+z=\frac{861.8\rm{KeV}}{E}}+\frac{0.104(1+z)^2}{384.2\rm{KeV}}\frac{dX_{\lis}^{decay}}{dz}|_{1+z=\frac{384.2\rm{KeV}}{E}}\right],
\end{equation}
where $n_{H0}$ is the density of hydrogen nuclei at $z=0$. The full width at
half maximum (FWHM) for the first line is $2.3$eV and the central energy is
$29.5$eV. For the second line the FWHM is $1$eV with a central energy of
$13.1$eV. The line width at half maximum is $\delta E/E =7.8\%$. The width and asymmetric line profile of these lines is defined by the
kinetics of recombination.

\section{Comparison with other sources of neutrinos and energy release before
  recombination}
At $T\sim 0.1\rm{MeV}$ annihilation of electrons and positrons makes the biggest
contribution to cosmic neutrinos, however these neutrinos  have a much
broader thermal spectrum and form part of the low energy cosmic neutrino
background (CNB). Decay of neutrons and other nuclear reactions during BBN
also contribute  neutrinos, but due to high redshift $z\sim 10^8$ these
neutrinos are also today redshifted to the sub-eV energies, $\sim
1\rm{MeV}/10^8\sim 0.01\rm{eV}$. The spectrum of these neutrinos would be
broad since they are emitted over a period when redshift changes by a
factor of $\sim 2$. Energy released
during BBN is also quickly thermalized and does not lead to spectral
distortions. Decay of tritium created during BBN occurs at $z\sim 2.5\times
10^5$ corresponding to half life of $12.32$ years. Decay of tritium to
helium-3 results in an electron with average energy of $Q_e=5.7$KeV and an
antineutrino with average energy $\sim 12.9$KeV. The antineutrino spectrum would
 be
broad  as they decay over a period corresponding to the age of the
Universe at that time and these antineutrinos would have an average energy
of $\sim 0.05 eV$ today assuming they have zero mass. About $10\%$ of tritium
has decayed by $z=6.35\times 10^5$ and $90\%$ by z=$1.35\times 10^5$
leading to the width of the neutrino spectrum of $\delta E/E\sim
1.4$. This broadening is comparable  to the intrinsic width of the neutrino spectrum
 $\delta E/E\sim E_{max}/E_{avg}\sim 18.6/12.9=1.4$.  Thus non-thermal neutrinos from  all  sources
 before
recombination other than $\bes$ decay would have a spectrum much broader than those from $\bes$
decay and would have  much lower energy. These low energy neutrinos would
 be non-relativistic today in  more than one  mass eigenstates \citep{pdg}
compared to relativistic neutrinos from $\bes$ decay.
The
energy released into the plasma from tritium decay in the form of energetic electrons would
result in a chemical potential of $\mu \sim (Q_e/T)(n_{{}^3H}/n_{\gamma}) \sim
10^{-15}$ in the CMB. This is about the same amount of entropy generated in
the beryllium decay much later.  

\section{Conclusions}
Lithium-7 observed today was originally produced as beryllium-7 during
primordial nucleosynthesis \citep{wfh67}. Although half life of beryllium atoms is very short,
it has to wait until the beginning of recombination epoch to decay. We have
calculated the exact redshift when this happens. We have also estimated the
effect of energy release during the decay on the cosmic microwave
background. In addition the neutrinos produced during the decay give rise
to unique narrow lines in the cosmic neutrino background. 
These lines are too weak to be observable today. We should mention that
the detection of the much more numerous but lower energy CNB
neutrinos is currently being discussed  \citep{neutrino}. The recombination and decay of
beryllium has nevertheless theoretical significance. It marks the end of
primordial nucleosynthesis and the beginning of recombination. 

\section*{Acknowledgements}
Rishi Khatri would like to thank Prof. Brian Fields for informing him about
beryllium decay in the early Universe and discussions on the subject.


\end{document}